\documentclass[pre,aps,12pt]{revtex4}
\usepackage{graphicx}

\begin{document}

\title{Ferrimagnetism in delta chain with anisotropic ferromagnetic and
antiferromagnetic interactions}
\author{D.~V.~Dmitriev}
\author{V.~Ya.~Krivnov}
\email{krivnov@deom.chph.ras.ru}
\affiliation{Institute of Biochemical Physics of RAS, Kosygin str. 4, 119334, Moscow,
Russia.}
\date{}

\begin{abstract}
We consider analytically and numerically an anisotropic
spin-$\frac{1}{2}$ delta-chain (sawtooth chain) in which exchange
interactions between apical and basal spins are ferromagnetic and
those between basal spins are antiferromagnetic. In the limit of
strong anisotropy of exchange interactions this model can be
considered as the Ising delta chain with macroscopic degenerate
ground state perturbed by transverse quantum fluctuations. These
perturbations lift the ground state degeneracy and the model
reduces to the basal XXZ spin chain in the magnetic field induced
by static apical spins. We show that the ground state of such
model is ferrimagnetic. The excitations of the model are formed by
ferrimagnetic domains separated by domain walls with a finite
energy. At low temperatures the system is effectively divided into
two independent subsystems, the apical subsystem described by the
Ising spin-$\frac{1}{2}$ chain and the basal subsystem described
by the XXZ chain with infinite $zz$ interactions.
\end{abstract}

\maketitle

\section{Introduction}

\begin{figure}[tbp]
\includegraphics[width=3in,angle=0]{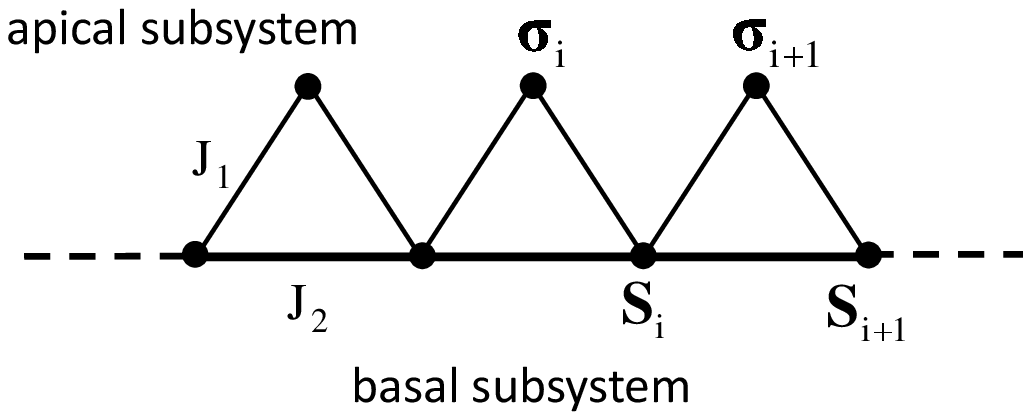}
\caption{The $\triangle $-chain model.}
\label{Fig_saw}
\end{figure}

The low-dimensional quantum magnets on geometrically frustrated lattices are
extensively studied during last years \cite{diep,mila}. An important class
of such systems is lattices consisting of triangles. An interesting and a
typical example of these objects is the $s=\frac{1}{2}$ delta or the
sawtooth Heisenberg model consisting of a linear chain of triangles as shown
in Fig.\ref{Fig_saw}. The interaction $J_{1}$ acts between the apical ($%
\sigma _{i}$) and the basal ($S_{i}$) spins, while $J_{2}$ is the
interaction between the neighboring basal spins. A direct interaction
between the apical spins is absent. The Hamiltonian of this model has a form
\begin{eqnarray}
\hat{H} &=&J_{1}\sum_{i=1}^{N}[S_{i}^{x}(\sigma _{i}^{x}+\sigma
_{i+1}^{x})+S_{i}^{y}(\sigma _{i}^{y}+\sigma _{i+1}^{y})+\Delta
_{1}S_{i}^{z}(\sigma _{i}^{z}+\sigma _{i+1}^{z})-\frac{\Delta _{1}}{2}]
\nonumber \\
&&+J_{2}\sum_{i=1}^{N}[S_{i}^{x}S_{i+1}^{x}+S_{i}^{y}S_{i+1}^{y}+\Delta
_{2}(S_{i}^{z}S_{i+1}^{z}-\frac{1}{4})]  \label{H1}
\end{eqnarray}%
where $\Delta _{1}$ and $\Delta _{2}$ are parameters representing the
anisotropy of the basal-apical and the basal-basal exchange interactions
respectively, $N$ is the number of triangles. The constants in this equation
are chosen so that the energy of the ferromagnetic state with the total spin
$L_{tot}^{z}=S_{tot}^{z}+\sigma _{tot}^{z}=\pm N$ is zero.

The isotropic delta chain ($\Delta _{1}=\Delta _{2}=1$) with both
antiferromagnetic interactions $J_{1}>0$ and $J_{2}>0$ (AF delta chain) has
been studied as a function of the parameter $\frac{J_{2}}{J_{1}}$ \cite%
{Sen,Nakamura,Blundell}. In spite of the simplicity of this model it
exhibits a variety of peculiar properties. If $\frac{J_{2}}{J_{1}}=1$ the
model has two-fold degenerate ground state where neighboring pairs of spins
form singlet configurations \cite{Nakamura}. When $\frac{J_{2}}{J_{1}}=\frac{%
1}{2}$ the delta chain supports the independent localized magnon states.
These states determine both the ground states properties and the
low-temperature thermodynamics in the vicinity of the saturation magnetic
field \cite{flat2008,Zhit,Schmidt,Honecker,Derzhko2004}. In particular, the
ground state is highly degenerate, the zero-temperature magnetization has a
plateau and the specific heat has the extra low-temperature peak.

In contrast to the AF delta chain the same model with $J_{1}<0$ and $J_{2}>0$
(the F-AF delta chain) is less studied. It is known \cite{Tonegawa} that the
ground state of the F-AF isotropic delta chain is ferromagnetic for $\alpha =%
\frac{J_{2}}{\left\vert J_{1}\right\vert }<\frac{1}{2}$. It was argued in
Ref.\cite{Tonegawa} on a base of numerical calculations that the ground
state for $\alpha >\frac{1}{2}$ is a special ferrimagnetic state. The
critical point $\alpha =\frac{1}{2}$ is the transition point between these
two ground state phases. The isotropic F-AF delta-chain at the transition
point $\alpha =\frac{1}{2}$ has been studied in Ref.\cite{KDNDR}. It was
shown \cite{KDNDR} that the ground state at the transition point (at zero
magnetic field) is macroscopically degenerate and consists of multi-magnon
configurations formed by independent localized magnons and the special
localized multi-magnon complexes.

The isotropic F-AF delta chain is a minimal model for the description of
several magnetic compounds such as malonato-bridged copper complexes of
formula $[Cu(bpy)H_{2}O]\times \lbrack Cu(bpy)(mal)H_{2}O](ClO_{4})_{2}$
containing magnetic $Cu^{2+}$ ions \cite{Inagaki,Tonegawa,ruiz,Kaburagi}.
From the analysis of the experimental data it was concluded \cite{Inagaki}
that the ratio of exchange interactions $\alpha =\frac{J_{2}}{\left\vert
J_{1}\right\vert }$ in this compound is $\alpha \simeq 1$. It means that
this compound is on the ferrimagnetic side of the ground state diagram of
the isotropic delta chain. Thus, the study of the ferrimagnetic state of the
F-AF delta chain is important and interesting problem. Numerical
calculations used in Ref.\cite{Tonegawa} suppose that the ground state
magnetization per site in the ferrimagnetic phase in the isotropic model is $%
\frac{1}{4}$. Unfortunately, numerical methods do not allow to obtain the
detail information about the structure and the properties of the
ferrimagnetic phase. At the same time this model is rather complicated and
can not be tractable analytically.

In this paper we show that the analysis of the anisotropic F-AF model in the
limit of high anisotropy helps to understand the origin and the properties
of the ferrimagnetic phase. For simplicity we consider the case of equal
basal-apical and the basal-basal anisotropy $\Delta _{1}=\Delta _{2}=\Delta $%
. In this case with $\Delta \gg 1$ the ferrimagnetic phase can exist in a
narrow interval of the value $\alpha $ (close to $\alpha =1$) between the
ferromagnetic (at $\alpha <\frac{\Delta }{1+\Delta }$) and the
antiferromagnetic (at $\alpha >1$) phases \cite{Kaburagi}. Therefore, in
order to investigate the ferrimagnetic phase we put $\alpha =1$. Then the
Hamiltonian of the F-AF delta chain can be represented in a form:
\begin{eqnarray}
\frac{1}{\Delta }\hat{H} &=&\frac{1}{\Delta }%
\sum_{i=1}^{N}(S_{i}^{x}S_{i+1}^{x}+S_{i}^{y}S_{i+1}^{y})-\frac{1}{\Delta }%
\sum_{i=1}^{N}[S_{i}^{x}(\sigma _{i}^{x}+\sigma _{i+1}^{x})+S_{i}^{y}(\sigma
_{i}^{y}+\sigma _{i+1}^{y})]  \label{1} \\
&&+\sum_{i=1}^{N}[S_{i}^{z}S_{i+1}^{z}-S_{i}^{z}(\sigma _{i}^{z}+\sigma
_{i+1}^{z})+\frac{1}{4}]  \nonumber
\end{eqnarray}%
where we put $J_{1}=-1$ and $J_{2}=1$.

The main aim of this paper is to study the model (\ref{1}) for $\Delta \gg 1$%
. We expect that some principal features of the ferrimagnetic phase of model
(\ref{1}) survive in the isotropic case.

Additional motivation of this study is related to the problem of
`order by disorder'. The fact is that the model (\ref{1}) in the
limit $\Delta \to \infty $ turns into the classical Ising model on
the delta chain with equal but opposite in sign apical-basal and
basal-basal interactions:
\begin{equation}
\hat{H}_{I}=\sum_{i=1}^{N}[S_{i}^{z}S_{i+1}^{z}-S_{i}^{z}(\sigma
_{i}^{z}+\sigma _{i+1}^{z})+\frac{1}{4}]  \label{Ising}
\end{equation}

It is known \cite{Sondhi,DK15} that the ground state of this model is
macroscopically degenerate and it is separated from the excited states by a
finite energy gap. This degenerate ground state is disordered (zero
magnetization), and the main question of the `order by disorder' problem is
what happens when such disordered system is perturbed by the quantum
fluctuations. The quantum fluctuations can lift the degeneracy and drive the
system to either ordered or disordered ground state. Generally, there are
many different ways of the introduction of such perturbations. One of them
is given by the transverse terms in Eq.(\ref{1}) and we will show that it
leads to the ordered ground state. On the contrary, the perturbation of the
Ising model (\ref{Ising}) by a transverse magnetic field results in the
disordered ground state \cite{Sondhi}.

Another example of influence of quantum dynamics on the Ising model (\ref%
{Ising}) was considered in Ref.\cite{DK15}, where the anisotropic
F-AF model (\ref{H1}) was studied for a special choice of the
exchange interactions and the anisotropies: $\alpha =1/(2\Delta
_{1})$ and $\Delta _{2}=(2\Delta _{1}^{2}-1)$. For such choice of
the interactions the F-AF model describes the phase boundary
between different ground state phases on the ($\alpha ,\Delta
_{1}$) plane and reduces to the Ising model (\ref{Ising}) at
$\Delta _{1}\to \infty $. The quantum fluctuations lift the ground
state degeneracy of Ising model (\ref{Ising}) but only partly, so
that the degeneracy remains macroscopic on this phase boundary, it
does not depend on
$\Delta _{1}$ and coincides with that for the isotropic F-AF delta-chain at $%
\alpha =\frac{1}{2}$. The spectrum of low-energy excitations has a highly
nontrivial multi-scale structure leading to the specific low-temperature
thermodynamics \cite{DK15}. This special model is another example of
`disorder by disorder' instead of `order by disorder'.

The paper is organized as follows. In Section II we study the spectrum of
model (\ref{1}) in different sectors of total spin $S_{tot}^{z}$ and show
that the ground state is ferrimagnetic one. In Section III we study the
low-temperature thermodynamics of the system both analytically and
numerically. In Section IV we give a summary of our results.

\section{Ferrimagnetic ground state}

At $\Delta \to \infty $ the model (\ref{1}) reduces to the Ising
model on the delta-chain described by Hamiltonian (\ref{Ising}).
The total $4^{N}$ eigenstates of this model is divided in two
subsets. The first one consists of degenerate ground states with
zero energy. These states include two types of the spin
configurations on triangles: either three spins in the triangle
have the same orientation or two basal spins of the triangle are
opposite oriented. In each triangle there are three configurations
which satisfy these conditions. Because the number of admissible
configurations is
the same for each triangle, the total number of the ground states is $3^{N}$%
. $(4^{N}-3^{N})$ states of the second subspace are separated from the
ground states by a `big' gap with the energy $E\sim 1$.

An infinitesimal perturbation of transverse interactions in Eq.(\ref{1})
lifts the macroscopic degeneracy of the ground state. However, a role of the
first and the second terms in lifting is different. The first term has
non-zero matrix elements both between the states of the first and the second
subsets while the second term in Eq.(\ref{1}) has non-zero matrix elements
between the states of the first and the second subsets only. Thus, only the
first term in Eq.(\ref{1}) gives contributions to an energy to the first
order in $\frac{1}{\Delta }$ whereas the second term is responsible for the
corrections which are proportional to $\frac{1}{\Delta ^{2}}$. Therefore, to
the leading order in $\frac{1}{\Delta }$ we can neglect the second term in
Eq.(\ref{1}) and the Hamiltonian (\ref{1}) reduces to that given by
\begin{equation}
\hat{H}=P[\Delta \hat{H}_{I}+%
\sum_{i=1}^{N}(S_{i}^{x}S_{i+1}^{x}+S_{i}^{y}S_{i+1}^{y})]P  \label{main}
\end{equation}%
where $P$ is a projector onto the first subspace containing
$3^{N}$ states and $\Delta \to \infty $ is assumed.

The model (\ref{main}) describes the basal $XXZ$ chain with infinite $zz$
interactions in the magnetic field produced by the static apical spins and
the magnetic field in the $i$-th basal site is $h_{i}=$ $\Delta (\sigma
_{i}^{z}+\sigma _{i+1}^{z})$. As a result, the magnetic field acting on the
basal spins depends on the spin configuration of apical subsystem. At first
we consider the most simple case when all apical spins are up (down)
producing the uniform magnetic field on basal subsystem: $h_{i}=\Delta $ ($%
h_{i}=-\Delta $). It is easy to check that if all apical spins are up
(down), the projector $P$ in Eq.(\ref{main}) eliminates the states in which
two basal spins down (two spins up) occupy neighboring sites. The total
number of allowable states is $(\frac{1+\sqrt{5}}{2})^{N}$ \cite{Domb}. The
Hamiltonian (\ref{main}) for the case $h_{i}=\Delta $ takes the form
\begin{equation}
\hat{H}=P_{0}\{\sum_{i=1}^{N}(S_{i}^{x}S_{i+1}^{x}+S_{i}^{y}S_{i+1}^{y})%
\}P_{0}  \label{oned}
\end{equation}%
where $P_{0}$ is the projector onto the states with no neighboring spins
down.

The model (\ref{oned}) can be mapped onto spinless fermions via the
Jordan-Wigner transformation
\begin{eqnarray}
S_{m}^{+} &=&c_{m}^{+}\exp (i\pi \sum_{l>m}c_{l}^{+}c_{l})\quad  \nonumber \\
S_{m}^{z} &=&\frac{1}{2}-c_{m}^{+}c_{m}  \label{JW}
\end{eqnarray}%
where $c_{m}^{+}$ is the Fermi-operator and we identify a spin down and a
spin up as a particle and a hole, correspondingly.

In fermion language the Hamiltonian (\ref{oned}) reads%
\begin{equation}
\hat{H}=P_{0}\{\frac{1}{2}\sum_{i=1}^{N}(c_{i}^{+}c_{i+1}+c_{i+1}^{+}c_{i})%
\}P_{0}  \label{HF}
\end{equation}%
and the projector $P_{0}$ forbids two particles to occupy neighboring sites.

The model of the spinless fermions with such constraint (infinite
nearest-neighbor interaction) can be mapped onto the model of
non-interacting fermions as follows \cite{Henley} (for simplicity,
we consider an open chain with $N$ sites). Each configuration of
$M$ fermions on $N$ sites with constraint is mapped to the
configuration of $M$ fermions on $(N-M+1)$ sites without
constraint by removing one empty site between two occupied sites.
The Hamiltonian of such model depends on a number of fermions and
has a form
\begin{equation}
\hat{H}(M)=\frac{1}{2}\sum_{i=1}^{N-M+1}(c_{i}^{+}c_{i+1}+c_{i+1}^{+}c_{i})
\label{HM}
\end{equation}

Besides, the matrix elements between the corresponding configurations of Eq.(%
\ref{HF}) and Eq.(\ref{HM}) are equal to each other. An equivalence of two
models means that the dispersion relation in the spin sector $S^{z}=\frac{N}{%
2}-M$ is
\begin{equation}
\varepsilon (k_{m})=-\cos k_{m}  \label{sp}
\end{equation}%
where
\begin{equation}
k_{m}=\frac{\pi m}{N-M+2}  \label{km}
\end{equation}%
with $m=1,2,\ldots N-M+1$.

According to Eq.(\ref{sp}) the ground state energy of model (\ref{HM}) in
the limit $N,M\gg 1$ but for a fixed fermion density $\rho =\frac{M}{N}$ is
\begin{equation}
E_{0}(\rho )=N\frac{1-\rho }{\pi }\sin \left( \frac{\pi \rho }{1-\rho }%
\right)  \label{Er}
\end{equation}

Minimization of $E_{0}(\rho )$ with respect to $\rho $ gives
\begin{equation}
\rho =\rho _{0}\simeq 0.3008  \label{rho0}
\end{equation}%
and
\begin{equation}
E_{0}(\rho _{0})\simeq -0.217N  \label{Erho0}
\end{equation}

Returning to the spin language, Eq.(\ref{rho0}) means that the ground state
of Eq.(\ref{oned}) is realized in the spin sector $S^{z}=N(\frac{1}{2}-\rho
_{0})$. Thus, the total spin of the ground state of delta chain (\ref{1}) is
\begin{equation}
L_{0}^{z}=N(1-\rho _{0})  \label{L0z}
\end{equation}

It follows from Eq.(\ref{Er}) that the energy of the lowest excitations in
this spin sector is
\begin{equation}
\varepsilon =\frac{\pi (1-\rho _{0})}{N}\sin \left( \frac{\pi \rho _{0}}{%
1-\rho _{0}}\right) ,  \label{epsilon}
\end{equation}%
i.e. the excitations are sound-like with the sound velocity
\begin{equation}
c=\sin (\frac{\pi \rho _{0}}{1-\rho _{0}})  \label{sound}
\end{equation}

The case with all apical spins down is considered in a similar way. In this
case the role of the Fermi-particles is played by the basal spins up and the
total ground state spin is $L_{z}^{0}=-N(1-\rho _{0})$.

We note that formulae similar to Eqs.(\ref{Er}) and (\ref{rho0})
have been obtained earlier by the Bethe-ansatz method
\cite{Alcaraz} in the problem of an asymmetric diffusion of
molecules with different size.

\begin{figure}[tbp]
\includegraphics[width=3in,angle=-90]{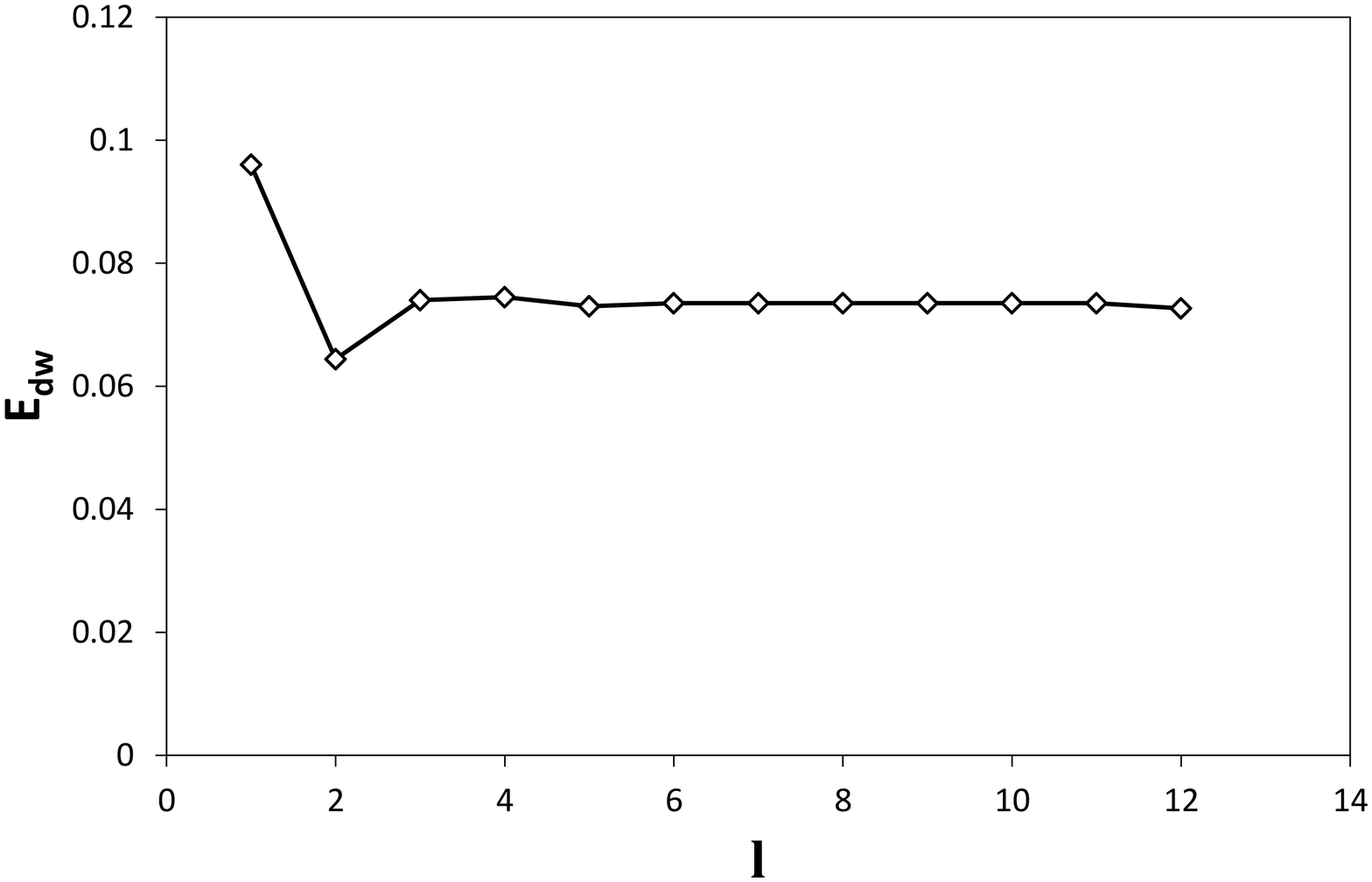}
\caption{Dependence of the domain wall energy on the domain size
$E_{\mathrm{dw}}(l)$ is calculated for the cyclic $XXZ$ chain of
length $N=24$ as a half of the energy difference between the
two-domain configuration with $l$ apical spins down and $(N-l)$
apical spins up and the one-domain ground state energy.}
\label{Edw-n}
\end{figure}

Eq.(\ref{Er}) with $\rho =\rho _{0}$ defines the ground state
energy of the Hamiltonian (\ref{main}) for the ferromagnetic
configuration of the apical subsystem. Now we need to consider
other distributions of up and down apical spins. This problem can
not be solved analytically and we use numerical calculations of
finite chains. These calculations show that the most important
configurations of the apical spin subsystem are the states with
alternating domains of the up and down spins. The simplest
configuration of such type is a two-domain structure consisting of
$l$ spins up and $(N-l)$ spins down separated by two domain walls
(for cyclic chains). For the two-domain configuration the magnetic
field induced by the apical spins is: $h=\Delta $ for $(l-1)$
basal sites; $h=-\Delta $ for $(N-l-1)$ sites; and $h=0$ on two
basal sites located in the center of two domain walls. (The
ferromagnetic state of the apical spins considered above
corresponds to $l=0$ or $l=N$ and it can be identified as the
one-domain structure). It is apparent that the minimal energy of
the two-domain state with $l,N\gg 1$ is reached when the density
of the fermions (in fermionic language) in each domain is $\rho
=\rho _{0}$. The total spin of this state is $L^{z}=(2l-N)(1-\rho
_{0})$. It is clear that the energy of this state is higher than
the ground state energy of the one-domain state due to the
presence of defects (the domain walls). The energy of the domain
wall $E_{\mathrm{dw}}(l)$ is defined as a half of the energy
difference between the two-domain configuration with $l$ apical
spins down and $(N-l)$ apical spins up and the one-domain ground
state energy. The numerical calculations on finite chain $N=24$
for the dependence of the domain wall energy on the domain size
$E_{\mathrm{dw}}(l)$ are shown in Fig.\ref{Edw-n}. The energies of
the one-domain and two-domain states are chosen for the optimal
value of the total $S^z$. As can be seen in Fig.\ref{Edw-n} the
domain wall energy $E_{\mathrm{dw}}$ slowly depends on $l$ when
the domain size $l\geq 2$ and $N\gg 1$ and rapidly converges to
the value $E_{\mathrm{dw}}\simeq 0.07$.

Similarly, any apical spin configuration can be represented as
many domain structure consisting of $r$ domains with spins up and
$r$ domains with spins down domains with $2r$ domain walls.
Numerical calculations show that the ground state energy of the
$r$-domain state is
\begin{equation}
E(r)=E_{0}+2rE_{\mathrm{dw}}  \label{E(r)}
\end{equation}%
where $E_{0}$ is the ground state energy of the one-domain
configuration ($r=0)$ given by Eq.(\ref{Er}).

In order to study the stability of the one-domain ground state
with respect to a creation of the two-domain states we consider
the dependence of the ground state of the one-domain configuration
with all apical spins up, $E_{0}(\rho )$, for $\rho $ close to
$\rho _{0}$. According to Eq.(\ref{Er}) the energy $E_{0}(\rho )$
has a minimum at $\rho _{0}$ and can be expanded in $\left\vert
\rho -\rho _{0}\right\vert \ll 1$ as
\begin{equation}
E_{0}(\rho )=E_{0}(\rho _{0})+bN(\rho -\rho _{0})^{2}  \label{12}
\end{equation}%
where%
\begin{equation}
b=\frac{\pi }{2(1-\rho _{0})^{3}}\sin \left( \frac{\pi \rho _{0}}{1-\rho _{0}%
}\right) \approx 4.46
\end{equation}

In an instability point the energies and the total spins of the
one- and two-domain states are equal. The total spins of the
one-domain state and two-domain one with $l$ up and $(N-l)$ down
apical spins are $L^{z}=N(1-\rho )$ and $L^{z}=(N-2l)(1-\rho
_{0})$, respectively. As a result the instability point is
determined by the relations
\begin{eqnarray}
bN(\rho -\rho _{0})^{2} &=&2E_{\mathrm{dw}}  \label{inst} \\
(\rho -\rho _{0}) &=&2(1-\frac{l}{N})(1-\rho _{0})  \nonumber
\end{eqnarray}

\begin{figure}[tbp]
\includegraphics[width=3in,angle=-90]{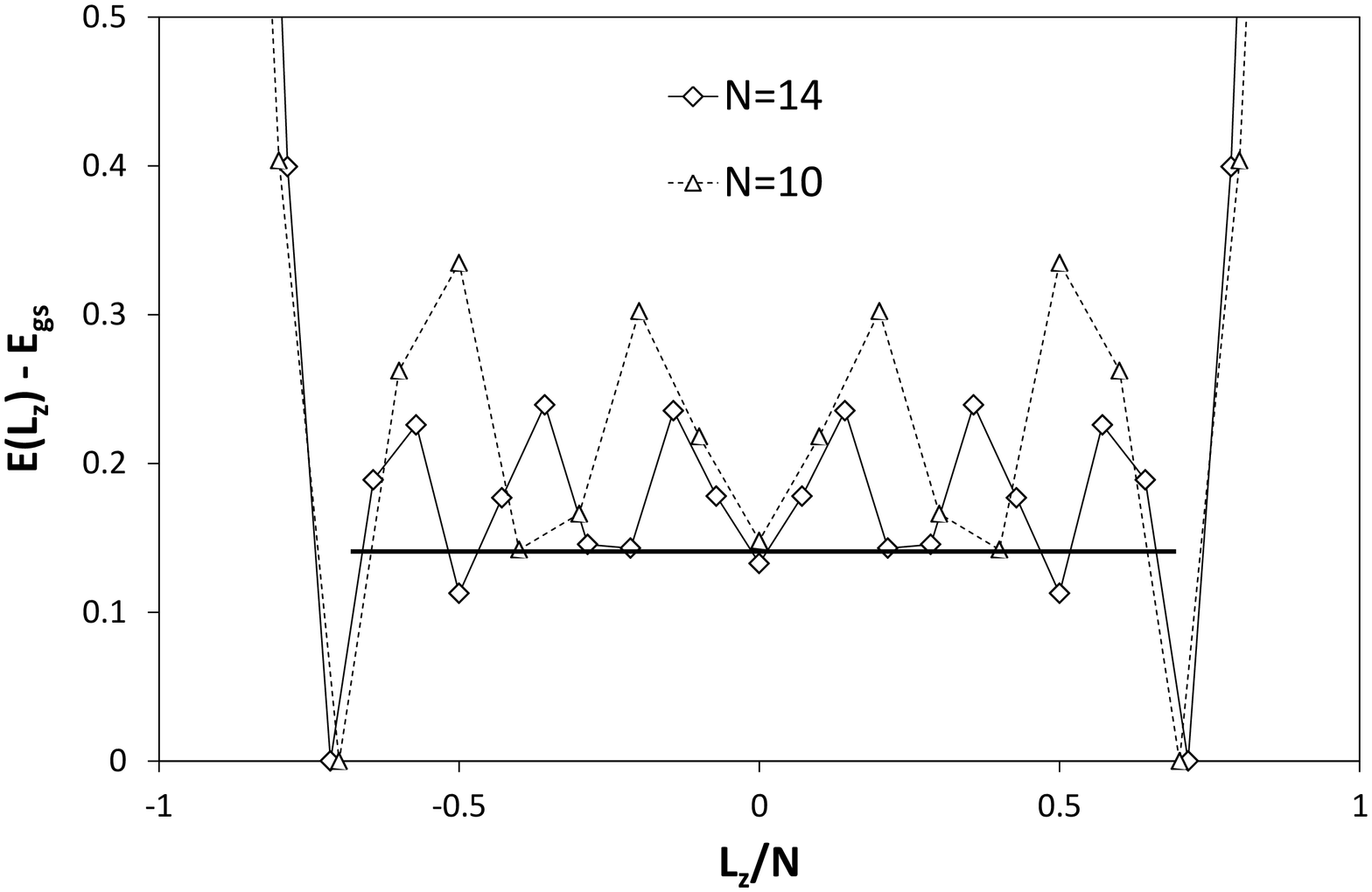}
\caption{Lowest energies in different sectors of total spin $L^{z}_{tot}$
for delta-chains with $N=10$ and $N=14$. Predicted thermodynamic limit is
shown by thick solid line.}
\label{Fig_E-Sz}
\end{figure}

As follows from Eqs.(\ref{inst}) the instability occurs for $\rho
>\rho _{0}$ and for small deviation from the minimum $(\rho-\rho_0)\sim N^{-1/2}$.
Thus, in the thermodynamic limit $N\to\infty $ the ground state is
realized for the one-domain state in the total spin sectors with
$|L^{z}|\geq L_{0}^{z}$ (see Eq.(\ref{L0z})), while in the sectors
$|L^{z}|<L_{0}^{z}$ the ground state corresponds to the two-domain
structure. But the global ground state of the model (\ref{main})
is twofold degenerate ferrimagnetic state with $L^{z}=\pm
L_{0}^{z}$. In these states the magnetization on apical and basal
sublattices are $\left\vert \left\langle \sigma
_{i}^{z}\right\rangle \right\vert =0.5$ and $\left\vert
\left\langle S_{i}^{z}\right\rangle \right\vert \simeq 0.2$, so
that the
total magnetization per site is $\left\vert \left\langle \frac{L^{z}}{2N}%
\right\rangle \right\vert =0.35$. The ground state energy as a function of $%
L^{z}/N$ obtained by numerical calculations of finite delta-chains
with $N=10 $ and $N=14$ is shown in Fig.\ref{Fig_E-Sz}. Irregular
form of this dependence is due to finite-size effects, which are
caused mainly by the deviation of the particle density $\rho =M/N$
possible for a given chain length $N$ from the optimal value $\rho
_{0}$. However, as it can be seen in Fig.\ref{Fig_E-Sz} the
amplitude of oscillations decreases with $N$ and the expected
thermodynamic limit $2E_{\mathrm{dw}}$ is shown in
Fig.\ref{Fig_E-Sz} by thick solid line.

\section{Low temperature thermodynamics}

The partition function $Z$ of the model (\ref{main}) is a sum of
contributions to $Z$ corresponding to all possible configurations of the
apical spins. Generally, each configuration of the cyclic delta-chain with $%
2r$ domain walls is specified by a set of $r$ domains of the apical spins up
with lengths $l_{1},l_{2},\ldots l_{r}$ and $r$ domains of the apical spins
down of length $m_{1},m_{2,}\ldots m_{r}$ which satisfy the conditions
\begin{equation}
\sum_{i=1}^{r}l_{i}=N-k,\quad \sum_{i=1}^{r}m_{i}=k  \label{rel}
\end{equation}%
where $k$ is a total number of down apical spins.

Then, the partition function $Z$ is
\begin{equation}
Z=\sum_{r}Z_{r}(l_{1},m_{1},l_{2},m_{2},\ldots l_{r},m_{r})  \label{Z}
\end{equation}%
where summation is carried out over $l_{i},m_{i}$ satisfying relations (\ref%
{rel}) and it includes two one-domain configurations with $r=0$.

\begin{figure}[tbp]
\includegraphics[width=3in,angle=-90]{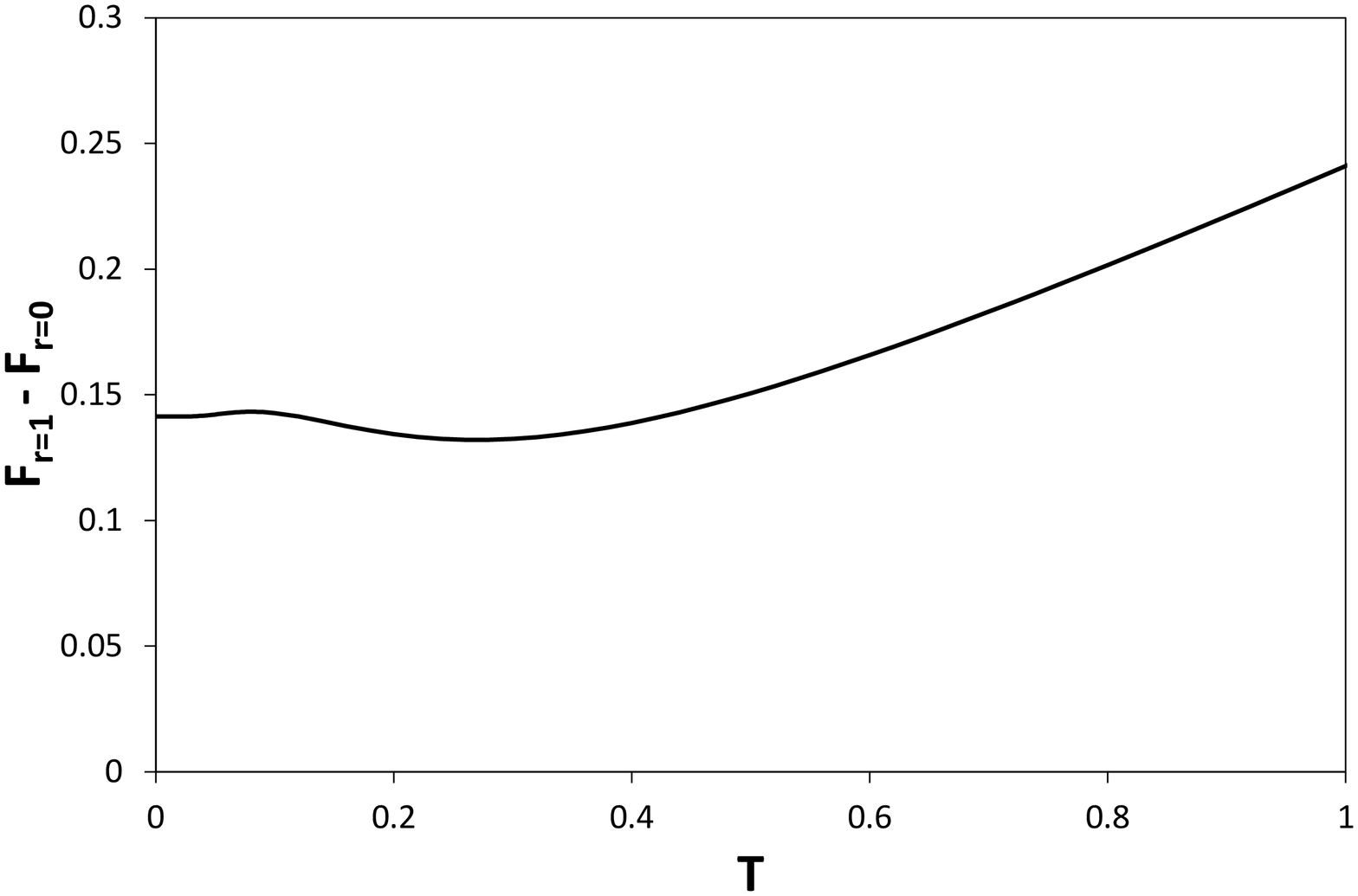}
\caption{Difference of the free energies of the two-domain ($r=1$)
and the one-domain ($r=0$) configurations as a function of $T$ for
the XXZ chain of length $N=16$.} \label{Fig_dF}
\end{figure}

The calculation of $Z$ in Eq.(\ref{Z}) is a complicated problem.
However, it can be simplified for low temperatures. As was noted
before the ground state energy of the configurations with $2r$
domain walls is higher than the one-domain state on the value
$2rE_{\mathrm{dw}}$. The same holds for the free energies. As an
example, we represent in Fig.\ref{Fig_dF} the difference between
the free energies of the one-domain ($r=0$) and two-domain ($r=1$)
configurations of cyclic chain with $N=8$ as the function of $T$.
This difference varies only slightly with $T$ and it is close to
the energy of two domain walls $2E_{\mathrm{dw}}$, so that the
deviation from the value $2E_{\mathrm{dw}}$ is less than $7\%$ for
$T<T_{1}\simeq 0.5$. It means that the two-domain partition
function $Z_{1}$ at $T<T_{1}$ can be written as
\begin{equation}
Z_{1}=Z_{0}\exp (-\frac{2E_{\mathrm{dw}}}{T})  \label{Z_1}
\end{equation}%
where $Z_{0}$ is the partition function of the model (\ref{oned}) describing
the one-domain configuration.

Similarly, if all domain sizes are large ($l_{i},m_{j}\gg 1$), the free
energy per site is the same for each domain and it is equal to that for the
one-domain configuration. Therefore, the partition function of the $r$ -
domain configuration can be approximately written as
\begin{equation}
Z_{r}=Z_{0}\exp (-2r\frac{E_{\mathrm{dw}}}{T})  \label{Zr}
\end{equation}

Then the partition function (\ref{Z}) takes the form%
\begin{equation}
Z=Z_{0}\sum_{r=0}^{N/2}\exp (-2r\frac{E_{\mathrm{dw}}}{T})W(r,N)  \label{Z0}
\end{equation}%
where $W(r,N)$ for $r\geq 1$ is the number of the configurations with $2r$
domain walls. The weights $W(r,N)$ are known \cite{Gaudin}
\begin{equation}
W(r,N)=\sum_{m=r}^{N-r}\frac{N}{m}C_{m}^{r}C_{N-m-1}^{r-1}  \label{W}
\end{equation}%
where $C_{n}^{k}$ are binomial coefficients and $W(0,N)=2$.

The sum in Eq.(\ref{Z0}) looks like the partition function of the 1D Ising
model of the apical spins $\sigma =\frac{1}{2}$ with the effective
nearest-neighbor ferromagnetic interaction $J=E_{\mathrm{dw}}$, i.e. the
partition function $Z$ at $T<T_{1}$ is a product of the partition functions
of the model (\ref{oned}) and that of the effective 1D Ising model $Z_{I}$,
i.e $Z=Z_{0}Z_{I}$. It means that the free energy and other thermodynamic
quantities are sums of those for the 1D Ising model and for the model (\ref%
{oned}). As to the thermodynamics of the latter it can be obtained using the
known spectrum of this model given by Eq.(\ref{sp}). Then, the free energy $%
F_{0}=-T\ln Z_{0}$ has a form
\begin{equation}
\frac{F_{0}}{N}=-\frac{T(1-\rho )}{\pi }\int_{0}^{\pi }\ln [1+\exp (\frac{%
\cos k+\mu }{T})]dk  \label{F0}
\end{equation}

The chemical potential $\mu $ and the density $\rho $ as functions of $T$
are determined from the equations $\partial F/\partial\rho=0$ and $\partial
F/\partial\mu=0$ with $F=F_0+\mu\rho$, which result in
\begin{eqnarray}
\mu &=&-\frac{T}{\pi }\int_{0}^{\pi }\ln [1+\exp (\frac{\cos k+\mu }{T})]dk
\nonumber \\
\rho &=&\frac{(1-\rho )}{\pi }\int_{0}^{\pi }[1+\exp (-\frac{\cos k+\mu }{T}%
)]^{-1}dk  \label{murho}
\end{eqnarray}

\begin{figure}[tbp]
\includegraphics[width=3in,angle=-90]{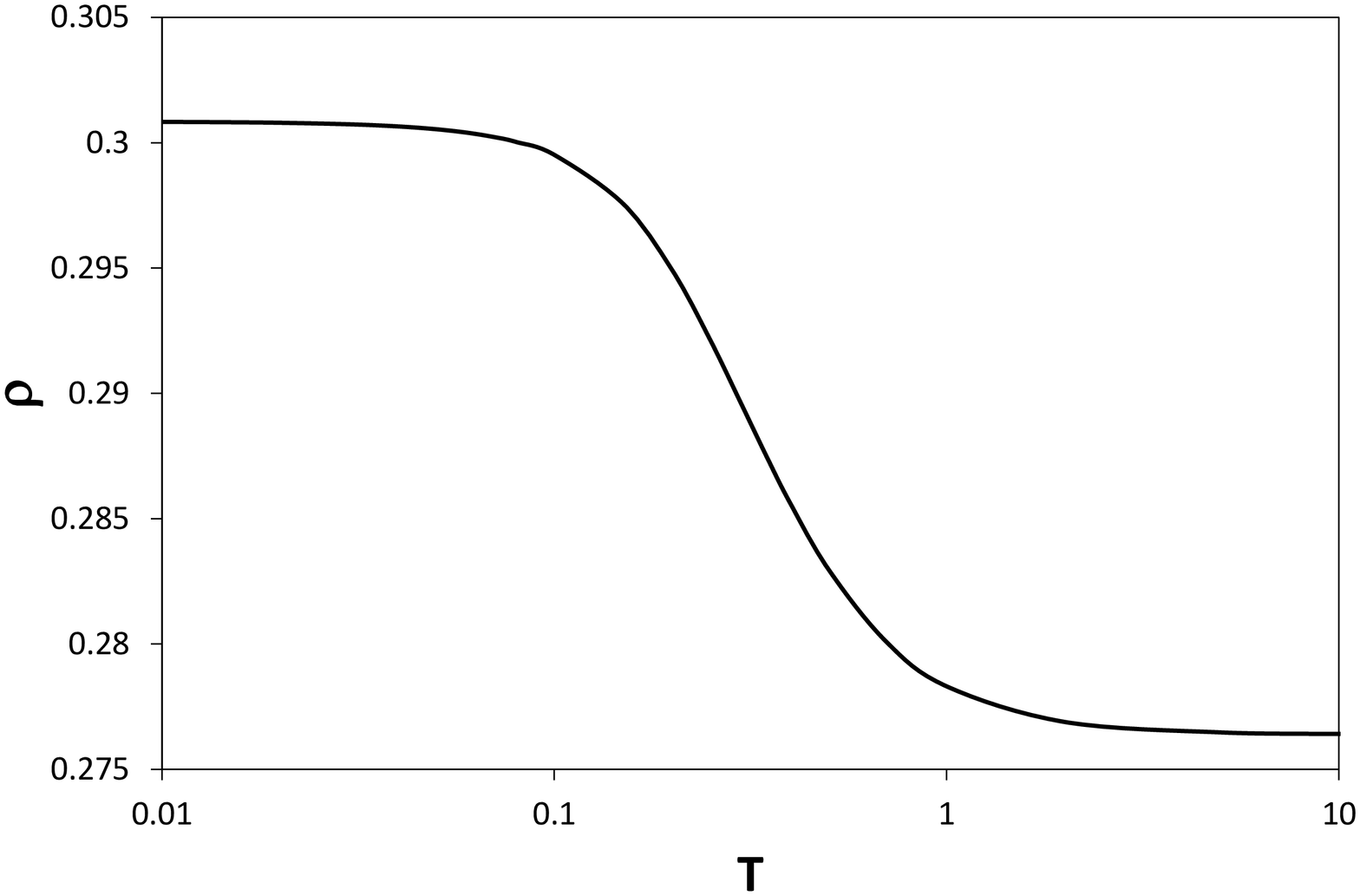}
\caption{Dependence $\rho(T)$.} \label{Fig_rho-T}
\end{figure}

In particular, the temperature dependence of the density $\rho (T)$ is shown
in Fig.\ref{Fig_rho-T}. As follows from Fig.\ref{Fig_rho-T} $\rho (T)$
changes from $\rho \simeq 0.3$ at $T=0$ to $\rho =(\sqrt{5}-1)/2\sqrt{5}%
\simeq 0.276$ at $T\gg 1$. The formula (\ref{F0}) coincides with that
obtained by different method in Ref.\cite{Klumper}, where the $XXZ$ chain in
the vicinity of the triple point has been studied.

\begin{figure}[tbp]
\includegraphics[width=3in,angle=-90]{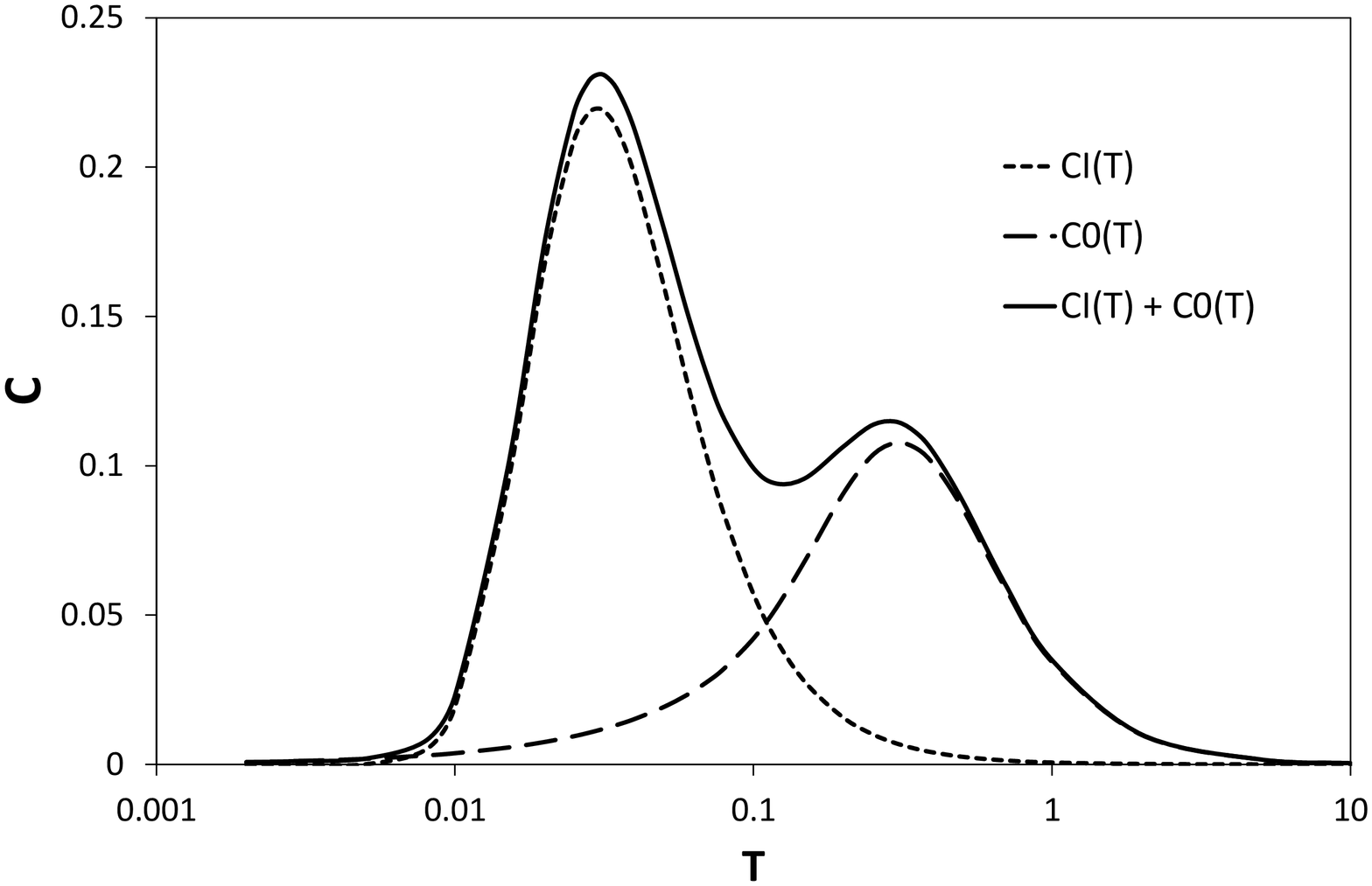}
\caption{Two contributions to the specific heat and their sum as a function
of $T$.}
\label{Fig_C-T_theory}
\end{figure}

Using Eq.(\ref{F0}) and well known thermodynamics of the 1D Ising model we
can obtain all thermodynamic quantities of the model (\ref{main}). As an
example, the specific heat $C(T)=C_{I}(T)+C_{0}(T)$ as a function of $T$ is
shown in Fig.\ref{Fig_C-T_theory} together with the contributions $C_{I}(T)$
and $C_{0}(T)$. The specific heat has a sharp maximum at $T\simeq 0.03$ and
the main contribution to it is given by the Ising term, while the shoulder
in $C(T)$ at $T\simeq 0.3$ is related to the maximum in $C_{0}(T)$. At $%
T\to 0$ the `Ising' contribution $C_{I}(T)$ is exponentially small
and the specific heat is uniquely determined by that for
$C_{0}(T)$
\begin{equation}
\frac{C}{N}=2(1-\rho _{0})\pi T\sin ^{-1}(\frac{\pi \rho _{0}}{1-\rho _{0}}%
),\quad T\to 0  \label{CT}
\end{equation}

As we noted before, Eqs. (\ref{Zr}) and (\ref{Z0}) are valid when the domain
sizes in the many domain configurations are large. To determine the
temperature region for which this is the case we use the steepest descent
method for the calculation of the sum in Eq.(\ref{Z0}). Using Stirling's
formula for the binomial coefficients in $W(r,N)$ we found that the main
contribution to the sum is given by the terms with
\begin{eqnarray}
k &=&\frac{N}{2}  \nonumber \\
r &=&\frac{N}{2}(1+\exp (\frac{E_{\mathrm{dw}}}{T}))^{-1}  \nonumber \\
l_{\downarrow } &=&l_{\uparrow }=(1+\exp (\frac{E_{\mathrm{dw}}}{T}))
\label{size}
\end{eqnarray}%
where $l_{\uparrow }$ and $l_{\downarrow }$ are average lengths $%
\left\langle l_{i}\right\rangle $ and $\left\langle m_{j}\right\rangle $ of
up- and down domains.

According to Eq.(\ref{size}) the representation of the partition function in
the form (\ref{Z0}) is valid if $\exp (E_{\mathrm{dw}}/T)\gg 1$ (or $T<E_{%
\mathrm{dw}}$). Because $E_{\mathrm{dw}}<T_{1}$ we conclude that the
partition function in the form (\ref{Z0}) secures a correct thermodynamics
of the model (\ref{main}) for $T<E_{\mathrm{dw}}$, while for $T>E_{\mathrm{dw%
}}$ it can give a qualitative description only.

Nevertheless, our calculations of finite systems show that all $r$ -
dependence of the partition function $Z_{r}(l_{1},m_{1},l_{2},m_{2},\ldots
l_{r},m_{r})$ for the configurations with small domains is expressed by the
factor $\exp (-2rE_{\mathrm{dw}}/T)$ where $E_{\mathrm{dw}}\simeq 0.07$ as
before. According to Eq.(\ref{size}) for $T>E_{\mathrm{dw}}$ the average
size of domains becomes $l_{\downarrow }=l_{\uparrow }\simeq 2$. Using these
facts we take as an approximation for the $r$ - domain partition function $%
Z_{r}(l_{1},m_{1},l_{2},m_{2},\ldots l_{r},m_{r})$ the expression in a form
\begin{equation}
Z_{r}=\tilde{Z}\exp (-2r\frac{E_{\mathrm{dw}}}{T})  \label{Z1}
\end{equation}%
where $\widetilde{Z}$ is the partition function for the up-up-down-down $%
(\uparrow \uparrow \downarrow \downarrow \uparrow \uparrow \downarrow
\downarrow \ldots )$ configuration of the apical spins.

\begin{figure}[tbp]
\includegraphics[width=3in,angle=-90]{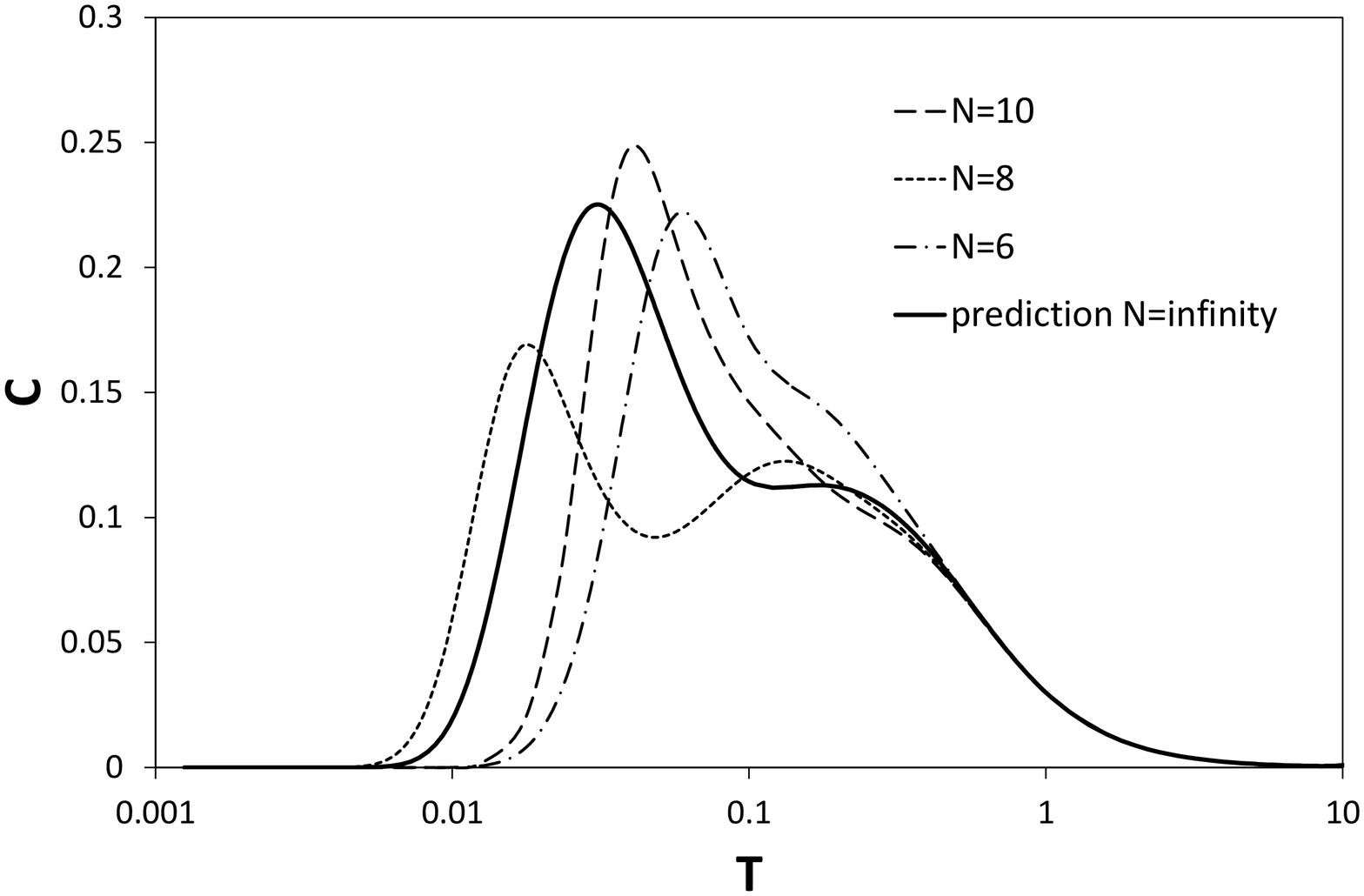}
\caption{Specific heat $C(T)$ calculated for delta chains with
$N=6,8,10$ and that predicted by approximation (\ref{ZZ}).}
\label{Fig_C-T}
\end{figure}

Then, the partition function $Z$ at $T>E_{\mathrm{dw}}$ is
\begin{equation}
Z=\tilde{Z}Z_{I}  \label{ZZ}
\end{equation}

\begin{figure}[tbp]
\includegraphics[width=3in,angle=-90]{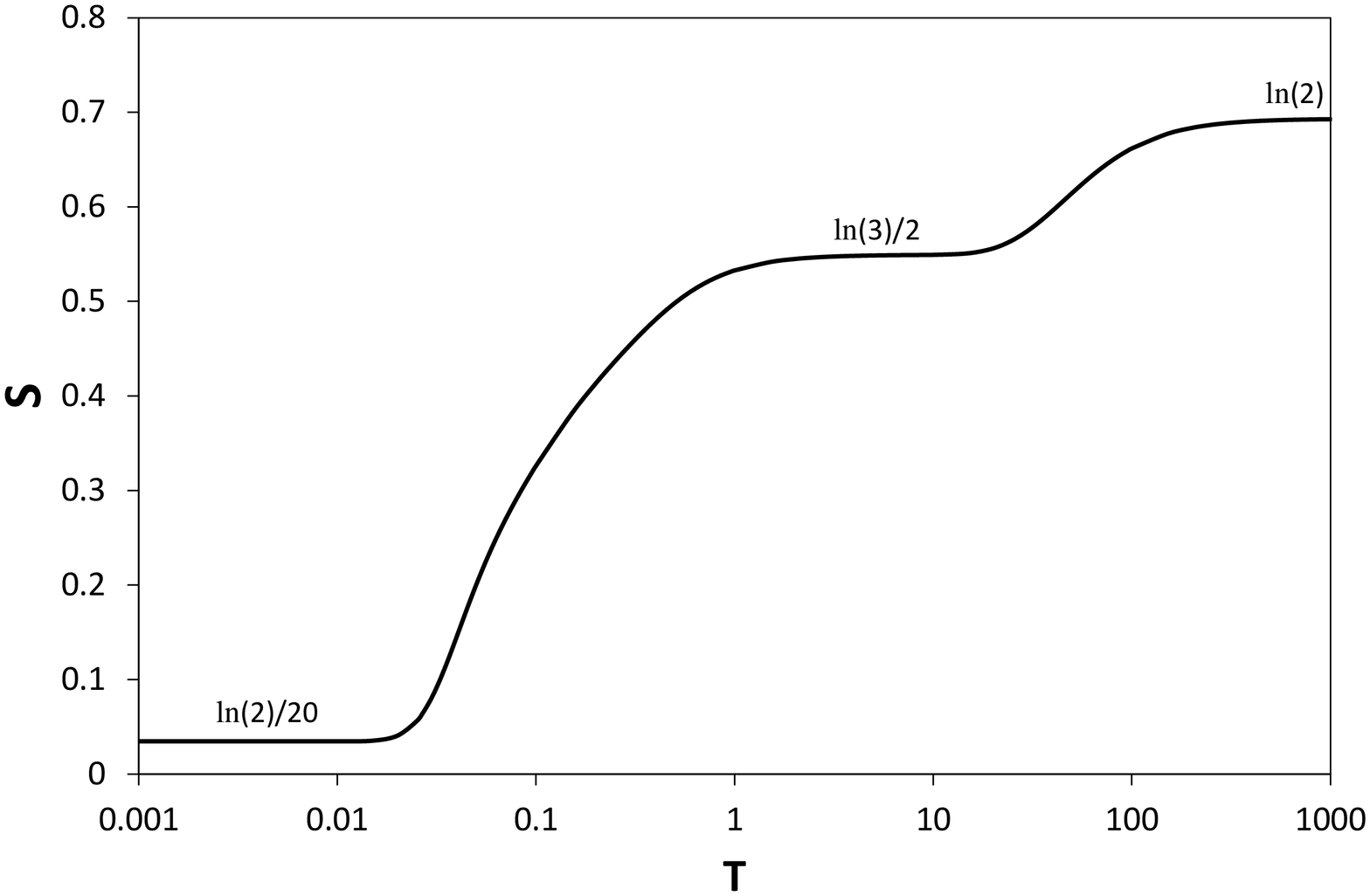}
\caption{Dependence of entropy per site on temperature $S(T)$ for
model (\ref{1}) with $\Delta =100$ and $N=10$.} \label{Fig_S-T}
\end{figure}

The thermodynamics of the up-up-down-down configuration is found by an exact
diagonalization (ED) calculation of finite chains. Corresponding results for
the specific heat are presented in Fig.\ref{Fig_C-T}. In Fig.\ref{Fig_C-T}
we also represent the results of the ED calculations of the model (\ref{1})
with $\Delta =100$ for $N=6,8,10$. We note that the model (\ref{1}) with
large but finite $\Delta $ and the model (\ref{main}) are formally
non-equivalent because the total number of states of these two models are
different and include $4^{N}$ and $3^{N}$ states, respectively. However, in
the temperature region $T<10$ the thermodynamics of the model (\ref{1}) is
governed by exactly $3^{N}$ states as follows from the temperature
dependence of the entropy per spin (see Fig.\ref{Fig_S-T}). Thus, at $T<10$
the thermodynamics of the models (\ref{main}) and (\ref{1}) is identical. As
it can be seen in Fig.\ref{Fig_C-T}, the  data for $C(T)$ for the model (\ref%
{1}) with different $N$ deviate at $T\lesssim E_{\mathrm{dw}}$ both from
each other and from the results for infinite system obtained from Eq.(\ref%
{Z0}). It means that the finite-size effects are essential in this
temperature region. On the other hand, the data for different $N$ are
indistinguishable at $T\gtrsim 1$, testifying that the finite-size data
correctly describe the thermodynamic limit. We note also that at $T\gtrsim 1$
these data are close to those obtained from Eq.(\ref{Z1}) for the
up-up-down-down configuration. At the same time, the thermodynamics based on
Eqs.(\ref{Z0}) show the qualitatively similar behavior of the specific heat
in this temperature region.

Lastly, we consider the temperature dependence of the zero-field
susceptibility $\chi (T)$. In this case it is necessary to include the
external magnetic field $h_{\mathrm{ext}}$ $\ll 1$ in the model (\ref{1}).
We confine ourself by the temperature region $T\lesssim E_{\mathrm{dw}}$
where the partition function is the product of the Ising and the one-domain
terms. We do not dwell on the technical details of the corresponding
computations. They are related to the solutions of Eqs.(\ref{F0}) and (\ref%
{murho}) as the functions of the temperature and the magnetic field $h_{%
\mathrm{ext}}$. The final result for the zero-field susceptibility $\chi (T)$
has the form
\begin{equation}
\frac{\chi (T)}{N}=2\chi _{I}(T)(1-\rho (T))+\chi _{0}(T)  \label{chi}
\end{equation}%
where $\chi _{I}(T)$ is the zero-field susceptibility per site of the
above-mentioned effective Ising model:
\begin{equation}
\chi _{I}=\frac{1}{4T}\exp (-\frac{E_{\mathrm{dw}}}{2T})  \label{chiI}
\end{equation}%
$\rho (T)$ is the solution of Eq.(\ref{murho}) with $h_{\mathrm{ext}}=0$ and
$\chi _{0}(T)$ is the susceptibility of the model (\ref{oned}) given by
\begin{equation}
\chi _{0}(T)=\frac{(1-\rho (T))^{3}}{\pi T}\int_{0}^{\pi }\exp (-\frac{\cos
k+\mu (T)}{T})[1+\exp (-\frac{\cos k+\mu (T)}{T})]^{-2}dk  \label{chi0}
\end{equation}%
with $\mu (T)$ determined by Eq.(\ref{murho}) with $h_{\mathrm{ext}}=0$.

The temperature dependence of the quantity $\chi (T)T$ is shown in Fig.\ref%
{Fig_chi-T}. The susceptibility $\chi _{I}$ is proportional to $\frac{1}{T}%
\exp (E_{\mathrm{dw}}/2T)$ at $T\to 0$ while $\chi _{0}(0)$ is
finite
\begin{equation}
\chi _{0}(0)=\frac{(1-\rho _{0})^{3}}{\pi \sin \left( \frac{\pi \rho _{0}}{%
1-\rho _{0}}\right) }  \label{chi00}
\end{equation}

\begin{figure}[tbp]
\includegraphics[width=3in,angle=-90]{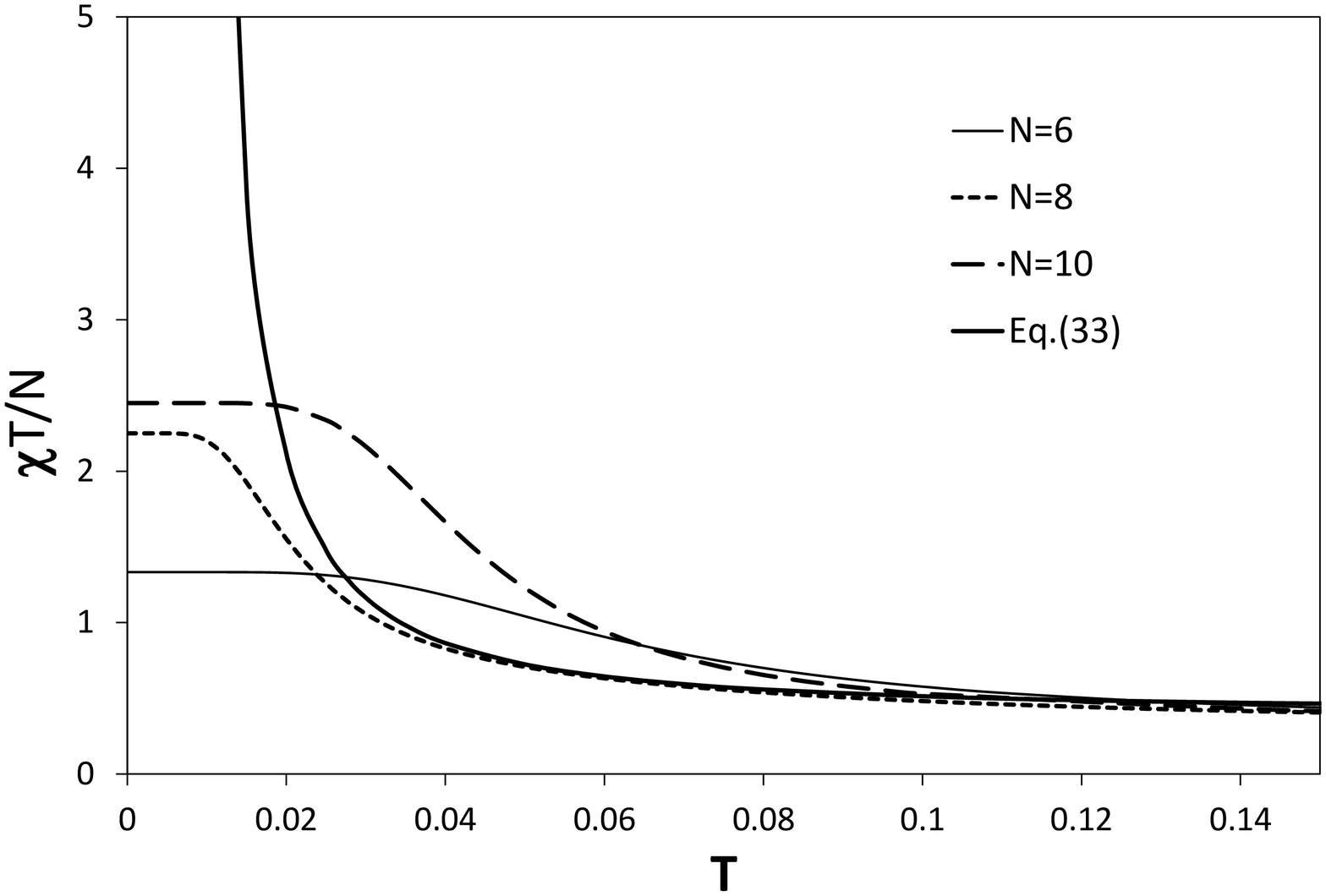}
\caption{Dependence of the susceptibility per site $\chi (T)T/N$
on $T$ for model (\ref{main}) with $N=6,8,10$. Analytical
prediction Eq.(\ref{chi}) is shown by thick solid line.}
\label{Fig_chi-T}
\end{figure}

Therefore, the behavior of the susceptibility at low temperatures
is determined by the `Ising' contribution $\chi _{I}$ and,
therefore, exponentially diverges at $T\to 0$. In
Fig.\ref{Fig_chi-T} we also represent the temperature dependence
of $\chi (T)T$ for finite delta-chains obtained by the ED
calculations of model (\ref{1}). In contrast to the analytics
predicting the exponentially divergence of $\chi (T)T$ in the
thermodynamic limit, the calculations of finite chains show the
finite limit for $\chi (T)T$ at $T=0$. Such behavior is related to
the fact that the value $\chi (T)T$ at $T=0$ for finite $N$ equals
to the square of the ground state spin which is
$L_{z}^{2}=N^{2}(1-\rho _{0})^{2}$, which turns into the
divergence of $\chi (T)T$ in the thermodynamic limit.

\section{Summary}

We have studied the spin-$\frac{1}{2}$ F-AF delta chain in the limit of
large anisotropy of exchange interactions. In this limit the model reduces
to the $1D$ $XXZ$ chain on basal sites in the static magnetic field
depending on the domain structure of the apical spins. The ground state is
twofold degenerate and magnetically ordered. In the ground state the apical
spins form a fully polarized state with $\left\vert \left\langle \sigma
_{i}^{z}\right\rangle \right\vert =0.5$ and the magnetization of the basal
spins is $\left\vert \left\langle S_{i}^{z}\right\rangle \right\vert \simeq
0.2$. Of particular interest are the excited states which involve the domain
walls separating the domains of one or another ground state. Based on the
domain statistics we reduced the low-temperature thermodynamics problem to
those for the effective 1D Ising model for the apical subsystem and the $1D$
$XXZ$ chain with infinite $zz$ interactions for the basal subsystem. The
correlation functions $\left\langle \sigma _{i}^{z}\sigma
_{i+r}^{z}\right\rangle $ and $\left\langle
S_{i}^{z}S_{i+r}^{z}\right\rangle $ behave similarly to 1D Ising ones with a
correlation length proportional to $\exp (E_{\mathrm{dw}}/2T)$ at low
temperatures.

This simple picture provides a starting point for the qualitative
understanding of the ferrimagnetic phase of the isotropic model. Preliminary
numerical results indicate that the ground state magnetization on the apical
and the basal sites does not change considerably when the anisotropy
parameter $\Delta $ decreases from the large value to $1$. In the isotropic
case they are $\left\langle \sigma _{i}^{z}\right\rangle =0.414$ and $%
\left\langle S_{i}^{z}\right\rangle =0.086$ \cite{Satoshi}. However,
additional symmetry of the isotropic model requires certain modifications of
the presented approach.

\begin{acknowledgments}
We would like to thank J.Richter for valuable comments on the
manuscript. The numerical calculations were carried out with use
of the ALPS libraries \cite{alps}.
\end{acknowledgments}

\end{document}